# Cryogenic characterization of 28nm FD-SOI ring oscillators with energy efficiency optimization

H. Bohuslavskyi, S. Barraud, V. Barral, M. Cassé, L. Le Guevel, L. Hutin, B. Bertrand, A. Crippa, X. Jehl, G. Pillonnet, L. Jansen, F. Arnaud, P. Galy, R. Maurand, S. De Franceschi, M. Sanquer, and M. Vinet

*Abstract*—Extensive electrical characterization of ring oscillators (ROs) made in high-κ metal gate 28nm Fully-Depleted Silicon-on-Insulator (FD-SOI) technology is presented for a set of temperatures between 296 and 4.3K. First, delay per stage ($\tau_P$), static current ($I_{STAT}$), and dynamic current ($I_{DYN}$) are analyzed for the case of the increase of threshold voltage ($V_{TH}$) observed at low temperature. Then, the same analysis is performed by compensating $V_{TH}$ to a constant, temperature independent value through forward body-biasing (FBB). Energy efficiency optimization is proposed for different supply voltages ($V_{DD}$) in order to find an optimal operating point combining both high RO frequencies and low power dissipation. We show that the Energy-Delay product (EDP) can be significantly reduced at low temperature by applying a forward body bias voltage ($V_{FBB}$). We demonstrate that outstanding performance of RO in terms of speed ($\tau_p$=37ps) and static power (7nA/stage) can be achieved at 4.3K with $V_{DD}$ reduced down to 0.325V.

## I. INTRODUCTION

Since the famous proposal for quantum computing with quantum dots [1], significant progress in Si spin qubits have been reported [2-5]. The first two-qubit logic gate in isotopically enriched Si was demonstrated [3] and a foundry-compatible CMOS SOI platform was used to demonstrate a hole spin qubit functionality [4-5]. These major achievements on the basic building block of a quantum computer are paving the way to the implementation of a large number of qubits individually controlled with tunable nearest-neighbor couplings. However, scaling up these systems in complex quantum computing architectures can be extremely challenging. In addition to fundamental advances in physical understanding and material development needed for spin qubits, a consistent engineering work should be made to propose a means of controlling, interacting and reading out a large number of qubits in parallel.

In recent years, hardware interfaces based on advanced CMOS technologies that operate at cryogenic temperature so as to ensure proximity to qubits have been proposed and discussed [6-9]. Bulk Si MOSFET [10-13] and other basic circuits such as Ring Oscillators (ROs) [11] and an FPGA (Field-Programmable Gate Array) [14] were characterized down to 4K. Despite a significant reduction of subthreshold swing (SS) and improvement of carrier mobility at low temperature, some limitations have been raised on bulk Si technologies. Non-ideal kink behavior and hysteresis in the characteristics are induced by the bulk current generated by impact ionization at the drain combined with increased resistivity from the freeze-out of charge carriers [15]. Moreover, the increase of drive current from the enhanced carrier mobility at low temperature is partially mitigated by the increase of threshold voltage which can be hardly compensated by back-biasing in bulk technologies.

In this work, an alternative to bulk Si CMOS technologies is proposed in order to provide more flexibility to designers for optimizing both high-performance and low power cryogenic electronics. Undoped thin-planar 28nm FD-SOI devices offer an excellent short-channel electrostatic control, low leakage current and immunity to random dopant fluctuations. More flexibility is brought to the circuit through an extremely effective back-biasing allowing to switch dynamically between high performance mode (Forward Body Biasing) and ultra-low leakage mode (Reverse Body Biasing) [16]. We thus propose to study the low temperature characterization of 28nm FD-SOI ring oscillators (RO) down to 4.3K. The RO performance in terms of delay per stage ($\tau_P$), dynamic ($I_{DYN}$), and static ($I_{STAT}$) currents is studied from 296 down to 4.3K for $V_{DD}$ ranging between 0.325 and 1.2V. Cryogenic effects on the RO performance are investigated with and without forward back-biasing (FBB) in order to compensate the shift of threshold voltage ($V_{TH}$) at low temperature. In addition, an energy efficiency optimization using FBB is proposed. We show that $V_{DD}$ can be reduced down to 0.325V while maintaining an ultra-low Energy-Delay product (EDP).

The paper is organized as follows: a brief review of 28nm FD-SOI is given in Section II. Low temperature characterization of FD-SOI ring oscillators and their energy efficiency optimization are discussed in Section III. Finally, the main conclusions are drawn in Section IV.

## II. BRIEF REVIEW OF 28NM FD-SOI TECHNOLOGY

The GO1 FD-SOI transistors are fabricated with a gate-first high-κ metal gate using STMicroelectronics technology [17-18]. They are processed on 300mm (100) SOI wafers with a buried oxide (BOX) thickness of 25nm. The equivalent oxide

H. Bohuslavskyi, S. Barraud, V. Barral, M. Cassé, L. Le Guevel, L. Hutin, G. Pillonnet, B. Bertrand and M. Vinet are with CEA, LETI, Minatec Campus, F-38054 Grenoble, France. X. Jehl, L. Jansen, A. Crippa, R. Maurand, S. De Franceschi, and M. Sanquer are with Univ. Grenoble Alpes, CEA, INAC-PHELIQS, F-38054 Grenoble, France. F. Arnaud and P. Galy are with STMicroelectronics, 850 rue J. Monnet, 38920 Crolles, France.
E-mail: sylvain.barraud@cea.fr

thickness is 1.55nm for n-MOS and 1.7nm for p-MOS. Low-threshold-voltage (LVT) transistors are used with the flip-well architecture: N-well (resp. P-well) with p-type (resp. n-type) back-plane doping for n-MOS (resp. p-MOS).

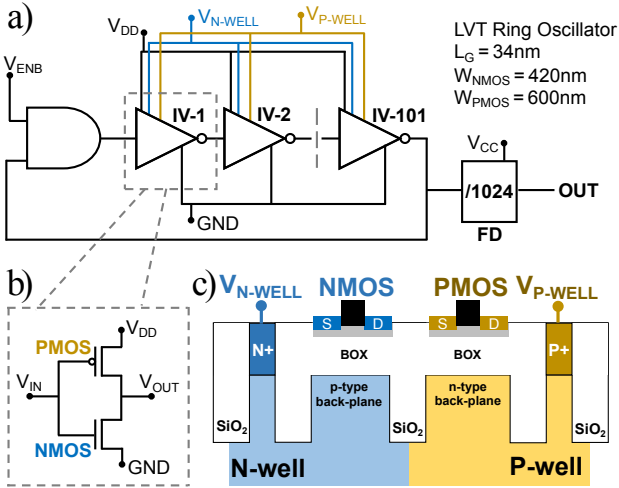

Fig. 1. (a) Schematic layout of 101-stages ring oscillator with 1024-frequency divider (FD). (b) Single inverter stage composed of n- and p-MOS. (c) Illustration of LVT transistors in the flip-well configuration. $V_{CC}$ is the supply voltage of the clock divider polarized with +1V. $V_{ENB}$ controls the AND gate that enables the oscillations ($V_{ENB}=V_{DD}$ was kept during all the measurements). For more details on the measurement protocol, see Table I.

TABLE I
RING OSCILLATOR AND MOSFET PARAMETERS, UNITS, DESCRIPTION, AND MEASUREMENT PROTOCOL

| Parameter | Unit | Description | Measurement protocol |
|---|---|---|---|
| $f$ | Hz | RO frequency | $V_{ENB}=V_{DD}$, $V_{CC}=1V$ and measure OUT. |
| $\tau_P$ | ps | RO delay per stage | $= 1/(f \times 2 \times 101 \times 1024)$. |
| $I_{DYN}$ | μA/stage | Dynamic current per stage in oscillating state | $V_{ENB}=V_{DD}$, $V_{CC}=$GND and measure OUT. |
| $I_{STAT}$ | nA/stage | Static current per stage in non-oscillating state | $V_{ENB}=V_{CC}=$GND and measure OUT. |
| $I_{EFF}$ | μA | Effective drive current of FD-SOI transistor [19] | $I_{EFF}=1/(1/I_{EFF-N}+1/I_{EFF-P})$ $I_{EFF-N(P)}=(I_H+I_L)/2$ where $I_H=I_{DS}(V_{GS}=V_{DD}, V_D=V_{DD}/2)$ $I_L=I_{DS}(V_{GS}=V_{DD}/2, V_D=V_{DD})$ where $V_{GS}$, $V_{DS}$, and $I_{DS}$ are MOSFET gate voltage, drain voltage and drain current. |
| $G(I_{EFF})$ | % | $I_{EFF}$ enhancement at low temperature | $= [I_{EFF}(T) - I_{EFF}(296K)] / I_{EFF}(T)$ |
| $G(\tau_P)$ | % | $\tau_P$ enhancement at low temperature | $= [\tau_P(T) - \tau_P(296K)] / \tau_P(T)$ |
| EPT | fJ | Energy per transition | $= 101 \times \tau_P \times (I_{DYN} - I_{STAT}) \times V_{DD}$ |
| EDP | fJ×ps | Energy Delay product | $= \tau_P \times$ EPT |

A 34nm gate length ($L_G$) is considered for both n- and p-MOS transistors of width $W_{NMOS}=420$nm and $W_{PMOS}=600$nm. The schematic layout of RO in the flip-well configuration is shown in Fig. 1. The RO consists of 101 identical stages together with an enabling two-way AND gate. The output is fed to a frequency divider to lower its frequency in the sub-MHz regime. The parameters and the measurement protocols used in this work are summarized in Table I.

### III. RESULTS AND DISCUSSION

#### A. RO performance at room temperature

In contrast to usual bulk technology, ultra-thin body and buried oxide FD-SOI enables an extended body-bias range from -3V (RBB) up to +3V (FBB), thanks to the thin buried oxide providing either high-performance or low-power transistors [20]. In order to compensate the increase of $V_{TH}$ at low-temperature this work is focused on LVT transistors which enable a strong improvement in the switching speed thanks to FBB (body factor of 85mV/V) at the cost of a higher leakage current. When no input is provided on the AND logic gate, $I_{STAT}$ and then static power consumption occur due to the flow of leakage from supply to ground. For the room temperature data in Fig. 2, the static current of 101-stages RO is plotted as a function of the delay per stage for different $V_{DD}$ and FBB voltages ($V_{FBB}$). Let us note that the delay of CMOS invertor can be approximated by $\tau_P = C_{LOAD} \times V_{DD}/I_{EFF}$ [21] with $I_{EFF}$ the effective current (defined in Table I) and $C_{LOAD}$ the load capacitance including the inversion capacitance, the parasitic capacitances and the wiring capacitance of back-end-of-line. Hence, lowering the supply voltage $V_{DD}$ to achieve lower static power dissipation (*i.e.* lower $I_{STAT}$) degrades the RO performance giving higher delay per stage. This frequency reduction can be balanced by using FBB voltage. The efficiency of forward back biasing for $\tau_P$ reduction and improved performance is well observed in Fig. 2. For the RO operating at $V_{DD}=1V$, a $\tau_p$ reduction of 33% is achieved (at $V_{FBB}=2.8V$) at the expense of a significant enhancement of static current (*i.e.* static power dissipation). Thus, a compromise in terms of power dissipation and speed should be carefully considered. Maintaining the back biasing efficiency at very low temperature is crucial to ensure accuracy and speed in the control and readout of qubits while using comparatively less power dissipation to perform the calculations.

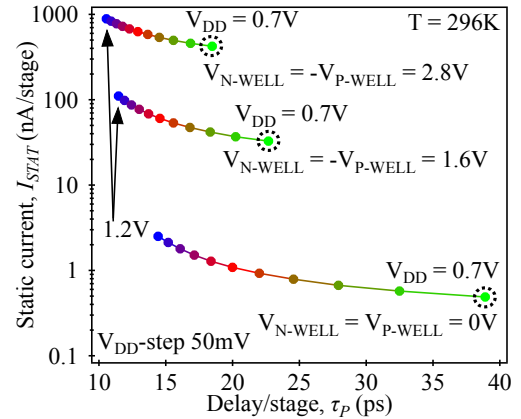

Fig. 2. Static current *vs* delay per stage for a set of $V_{DD}$ from 0.7 to 1.2V ($\Delta V_{DD}=50$mV). Three different FBB voltages are considered. $I_{STAT}$ is significantly increased at high $V_{FBB}$ illustrating that ROs are well optimized at room temperature with $V_{FBB}=0V$.

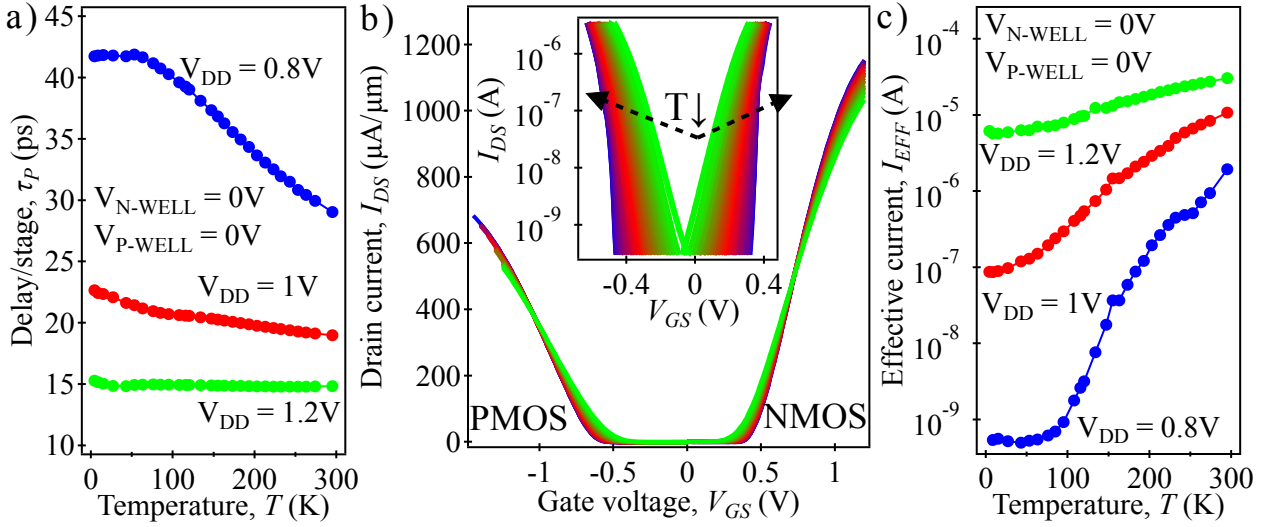

Fig. 3. (a) Delay per stage *vs* temperature for $V_{DD}=0.8$, 1, and 1.2V showing the RO slowing down due to the increase of $V_{TH}$ at low temperature. (b) $I_{DS}$-$V_{GS}$ curves recorded at $V_{DD}=1$V plotted in linear scale. The insert shows the subthreshold regime where current is plotted in logarithmic scale. Green (resp. blue) color corresponds to room temperature (resp. 4.3K). (c) Effective current *vs* temperature for different $V_{DD}$. The procedure used to calculate $I_{EFF}$ is described in Table I.

### B. RO performance without FBB down to 4.3K

Operation of FD-SOI transistors at cryogenic temperature was already reported in [18,22,23]. Since the scattering of charge carriers with phonons is sufficiently weak and can be neglected at liquid helium temperature, electron and hole mobilities are enhanced and should lead to a smaller $\tau_p$ at lower temperature for a given $V_{DD}$. However, despite a significant increase of drive current expected at low temperature [18], the RO slows down as it can be seen in Fig. 3a. This increase of delay per stage is explained by the $V_{TH}$-shift at lower temperatures (Fig. 3b) for both n- and p-MOS transistors. Consequently, the effective current is strongly reduced at lower temperatures due to a lower overdrive current for the gate voltage. This $I_{EFF}$ reduction becomes especially important at low $V_{DD}$. Hence, without FBB ($V_{N-WELL}=V_{P-WELL}=0$V) the enhanced carrier mobility at low temperature does not improve the effective current which becomes strongly limited by the $V_{TH}$ increase. The $I_{EFF}$ reduction shown on Fig. 3c is then directly responsible for the observed increase of $\tau_P$. The degradation of RO performance becomes even more important at low $V_{DD}$. Therefore, without forward-back-biasing used to compensate the $V_{TH}$-shift it may be difficult to work with an optimized cryogenic digital control electronics combining the demands of both high-performance and low-power consumption.

### C. RO performance by applying FBB down to 4.3K

In order to preserve the benefit of higher carrier mobility and thus, higher driving current at low temperatures, the $V_{TH}$-shift should be compensated. The ability to adjust $V_{TH}$ through back-biasing was already successfully demonstrated down to 4.3K using 28nm FD-SOI transistors [18]. In this work, the threshold voltages of n- and p-MOSFET as well as the body-factors ($\Delta V_{TH}/\Delta V_{FBB}$) were systematically extracted over a wide range of $V_{FBB}$ from 0 up to 3V. Then, by adjusting $V_{N-WELL}$ and $V_{P-WELL}$ for each temperature, the $V_{TH}$-shift was compensated (in order to keep the $V_{TH}$ measured at 300K) and thus, $V_{TH\_NMOS}$ and $V_{TH\_PMOS}$ were kept constant in temperature as shown in Fig. 4b. In contrast to previous results (without FBB), a significant speed up for different $V_{DD}$ is now observed when threshold voltages were settled to $V_{TH}(T)=V_{TH}(296K)$ within the full temperature range (Fig. 4a). By reducing the temperature from room temperature down to 4.3K, the delay per stage is decreased by 38% at $V_{DD}=0.8$V, 33% at $V_{DD}=1$V, and 31% at $V_{DD}=1.2$V. The $\tau_P$ reduction can be explained by the increase of $I_{EFF}$ shown in Fig. 4c. By monitoring the enhancement of $I_{EFF}$ and $\tau_P$ at lower temperatures (down to 4.3K) we observe a good correlation between the relative variations of $I_{EFF}$ increase and the $\tau_P$ reduction for $V_{DD}=0.8$V as shown in Fig. 5. Hence, the load capacitance seems to be weakly dependent on temperature since the $I_{EFF}$ increase is mainly responsible for the speed up of RO at low temperature.

Besides the low temperature behavior of $\tau_P$, other critical parameters such as dynamic current ($I_{DYN}$) and static current ($I_{STAT}$) were monitored for the evaluation of the power dissipation in oscillating and sleeping mode. As shown in Fig. 6, $I_{DYN}$ naturally increases as the temperature is reduced due to the $I_{EFF}$ enhancement, which leads to the increase of active power dissipation. Indeed, by applying FBB for the $V_{TH}$ compensation, the effective current is improved due to higher carrier mobility at low temperature and then $I_{DYN}$ is enhanced. In comparison to the room temperature case, the $I_{DYN}$ enhancement normalized by $V_{DD}$ is about 34% at 4.3K. At the same time a significant reduction in static current (Fig. 6b) and thus, in static power dissipation is observed thanks to the decrease of subthreshold slope at low temperature [18].

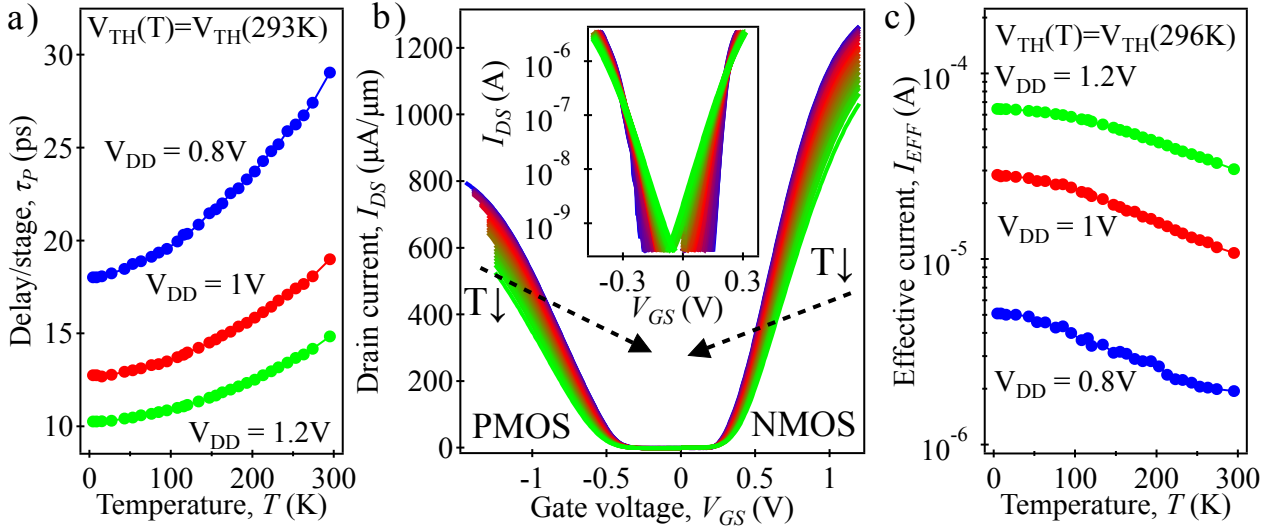

Fig. 4. (a) Delay per stage *vs* temperature for $V_{DD}$ = 0.8V, 1V, and 1.2V in the case of compensated $V_{TH}$. The RO speeds up at low temperature due to the carrier mobility enhancement. (b) $I_{DS}$-$V_{GS}$ curves recorded at $V_{DD}$=1V plotted in linear scale. The insert shows the subthreshold behavior with current plotted in logarithmic scale. $V_{TH\_NMOS}$ and $V_{TH\_PMOS}$ within the whole temperature range are constant and can be found in Table II. Green (resp. blue) color corresponds to room temperature (resp. 4.3K) (c) Effective current *vs* temperature for different $V_{DD}$. As it can be seen in Fig. 4b, the effective current increases at any stage of cooling down.

As compared to room temperature, the calculated static power dissipation $P_{STAT}=I_{STAT} \times V_{DD}$ at 4.3K is reduced by a factor of 1600 ($V_{DD}$=0.8V), 100 ($V_{DD}$=1V), and 6.5 ($V_{DD}$=1.2V). The RO performance in terms of $\tau_P$, $I_{DYN}$, and $I_{STAT}$ measured at 4.3 and 296K with and without FBB are summarized in Table II.

### D. Energy efficiency optimization down to 4.3K.

Wiring up large qubit arrays today is a key issue across all qubit platforms. Indeed, a large number of qubits must be read out periodically and rapidly processed to check whether errors occur along the way and to correct them. Furthermore, the energy efficiency of a cryogenic platform for the classical control of a scalable quantum computer is essential. Low power dissipation is required, down to few watts at 4K [24] to enable

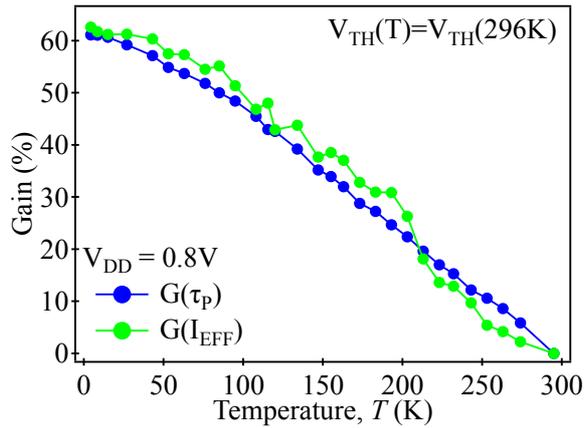

Fig. 5. Relative enhancement of $I_{EFF}$ and $\tau_p$ at $V_{DD}$ = 0.8V for temperature between 296 and 4.3K. FBB is applied to compensate the increase of $V_{TH}$ at low temperatures. Note that both relative gains are well correlated within the whole temperature range.

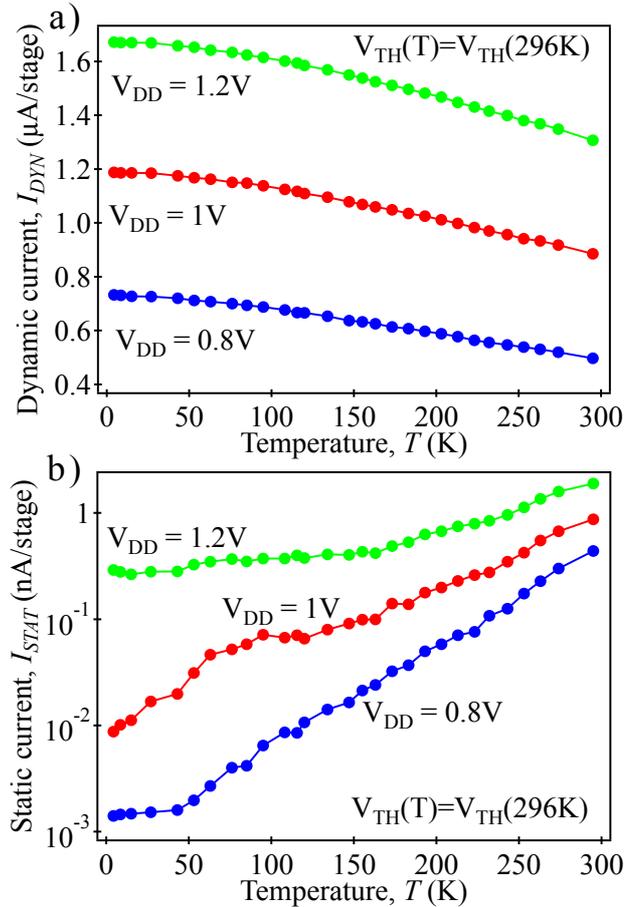

Fig. 6. (a) Dynamic current and (b) static current as a function of temperature for different $V_{DD}$ in case of compensated $V_{TH}$. Note that $I_{STAT}$ continuously decreases and reaches 2.3 pA/stage for $V_{DD}$=0.8V at 4.3K.

TABLE II
COMPARISON OF RING OSCILLATOR PERFORMANCE AT 296K AND 4.3K BETWEEN UNCOMPENSATED AND COMPENSATED $V_{TH}$

| $T$ (K) | $V_{DD}$ (V) | $V_{TH\_NMOS}$ (V) / $V_{TH\_PMOS}$ (V) | $V_{N\text{-}WELL}$ (V) / $V_{P\text{-}WELL}$ (V) | $\tau_P$ (ps) | $I_{DYN}$ (nA/stage) | $I_{STAT}$ (nA/stage) |
|---|---|---|---|---|---|---|
| 296 | 0.8 | 0.224 / -0.376 | 0 / 0 | 27 | 497 | 0.67 |
| | 1 | 0.216 / -0.363 | 0 / 0 | 18 | 885 | 1.29 |
| | 1.2 | 0.208 / -0.352 | 0 / 0 | 14 | 1308 | 1.9 |
| 4.3 | 0.8 | 0.391 / -0.549 | 0 / 0 | 42 | 363 | 0.002 |
| | | 0.224 / -0.376 | 1.65 / -2.75 | 18 | 717 | 0.002 |
| | 1 | 0.384 / -0.531 | 0 / 0 | 22 | 812 | 0.007 |
| | | 0.216 / -0.363 | 1.65 / -2.65 | 13 | 1182 | 0.01 |
| | 1.2 | 0.375 / -0.518 | 0 / 0 | 15 | 1281 | 0.037 |
| | | 0.208 / -0.352 | 1.62 / -2.62 | 10 | 1669 | 0.29 |

The set of $V_{N\text{-}WELL}$ and $V_{P\text{-}WELL}$ used for different $V_{DD}$ is chosen to compensate the corresponding $V_{TH}$-shift at 4.3K.

the operation of thousands of cryogenic fault-tolerant loops in existing refrigerators. The adoption of FD-SOI technology to reduce power consumption while keeping high speed of operation can be a promising solution in future implementations of large-scale quantum computers. Then, ensuring that control signals produce minimal dissipation will be essential at the lowest temperature stage of the refrigerator. We thus propose to deal with the energy consumption per transition and more specifically with the Energy-Delay metric to translate the more and more stringent constraint on the speed, while not disregarding the energy dissipation. The temperature dependence of Energy per Transition (EPT) for different $V_{DD}$ is shown in Fig. 7a. As in the previous section, FBB is applied to correct the $V_{TH}$-shift at low temperatures. One can notice that despite an increase of $I_{DYN}$ (shown in Fig. 6a), the significant decrease of delay per stage at low temperature (Fig. 4a) results in EPT becoming smaller and smaller during the cooling down to 4.3K. Lowering $V_{DD}$ at liquid helium temperature also leads to a significant reduction of EPT at the expense of performance with a lower RO frequency. To achieve the best possible performance, the Energy-Delay metric [21, 25] defined in Table I can be used. Smaller Energy-Delay values imply a lower energy consumption at the same level of performance corresponding to a more energy-efficient design. At RT, the lowest EDP is obtained for $V_{DD}$=1V as shown on Fig. 7b. However, as the RO is cooled down, we see that the optimal $V_{DD}$ can be lowered. At 4.3K, the minimal EDP is now obtained at $V_{DD}$=0.8V. The $V_{DD}$ dependence on the Energy-Delay Product is shown in Fig. 8a at 4.3 and 296K. At 4.3K, FBB is first used to keep the same $V_{TH}$ as that obtained at room temperature. Then, higher $V_{FBB}$ was applied in order to maximize the energy efficiency. As expected, EDP is minimal for $V_{DD}$=1V at 296K. However, a strong increase is observed at low $V_{DD}$ due to the rise of $\tau_P$ as shown on Fig. 3a. At 4.3K, the $V_{TH}$-compensation allows to reduce $\tau_P$ (see Fig. 4a) and leads to the decrease of EDP with a minimal value close to $V_{DD}$=0.8V. However, once again, the EDP remains strongly enhanced at low $V_{DD}$. The best FBB configuration minimizing EPT while keeping high RO speed is obtained for $V_{N\text{-}WELL}$=4V and $V_{P\text{-}WELL}$=−5.8V. This FBB configuration was chosen such as no minimum is observed in EDP even at $V_{DD}$ as low as 0.325V. The combination of LVT transistors with high forward-back-biasing allows to further reduce $V_{TH}$ and then to obtain lower $\tau_P$ for a given $V_{DD}$. In Fig. 8b, we show the static current versus $\tau_p$ for different $V_{DD}$ in order to highlight the advantage of FBB on LVT FD-SOI transistors. At 4.3K, low delay operation can be achieved without suffering from excessive increase of static power. However, if high FBB voltages ($V_{N\text{-}WELL}$ = 4V and $V_{P\text{-}WELL}$= −5.8V) are applied to reduce $V_{TH}$ close to 0V, $I_{STAT}$ is significantly increased. Nevertheless, excellent RO performance with $\tau_P$=37ps and $I_{STAT}$=7nA/stage are demonstrated at $V_{DD}$ = 0.325V. This static current can be further reduced by a factor 10 if $V_{P\text{-}WELL}$ is lowered from −5.8 to −4.8V while keeping $V_{N\text{-}WELL}$=4V. In addition, as shown in Fig. 8c, ultra-low $I_{DYN}$ current of 92nA/stage (at $V_{DD}$=0.325V) can be of great importance if low power dissipation in oscillating mode at moderate speed ($\tau_P$=71ps) is required. It should be noted that the RO performance for different FBB configurations can be further optimized depending on targeted applications.

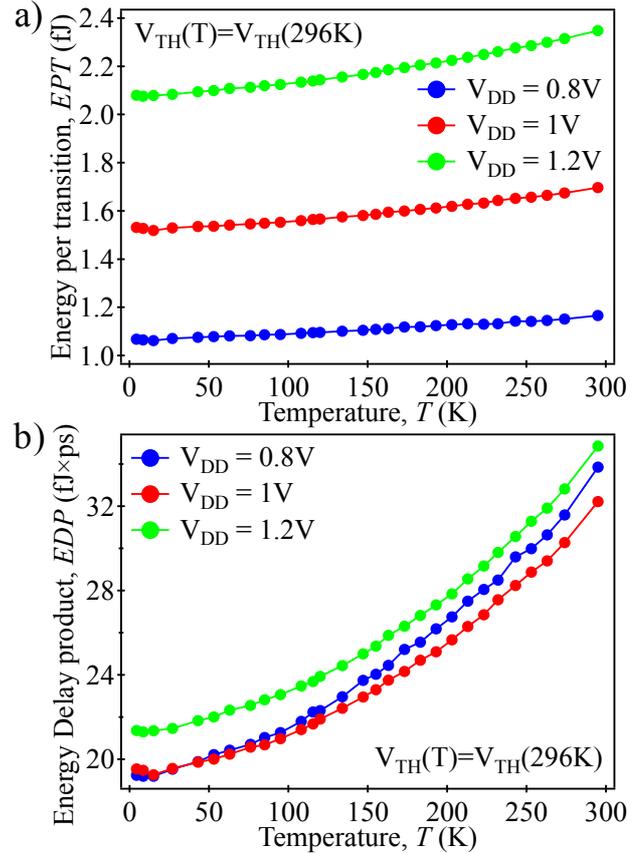

Fig. 7. (a) Energy per transition vs temperature for three different $V_{DD}$. Note that the decrease of EPT for all $V_{DD}$ at cryogenic temperature is mainly due to an important $\tau_P$ reduction. (b) Energy-Delay product vs temperature for $V_{DD}$=0.8, 1, and 1.2V. The lowest EDP indicates the optimal $V_{DD}$.

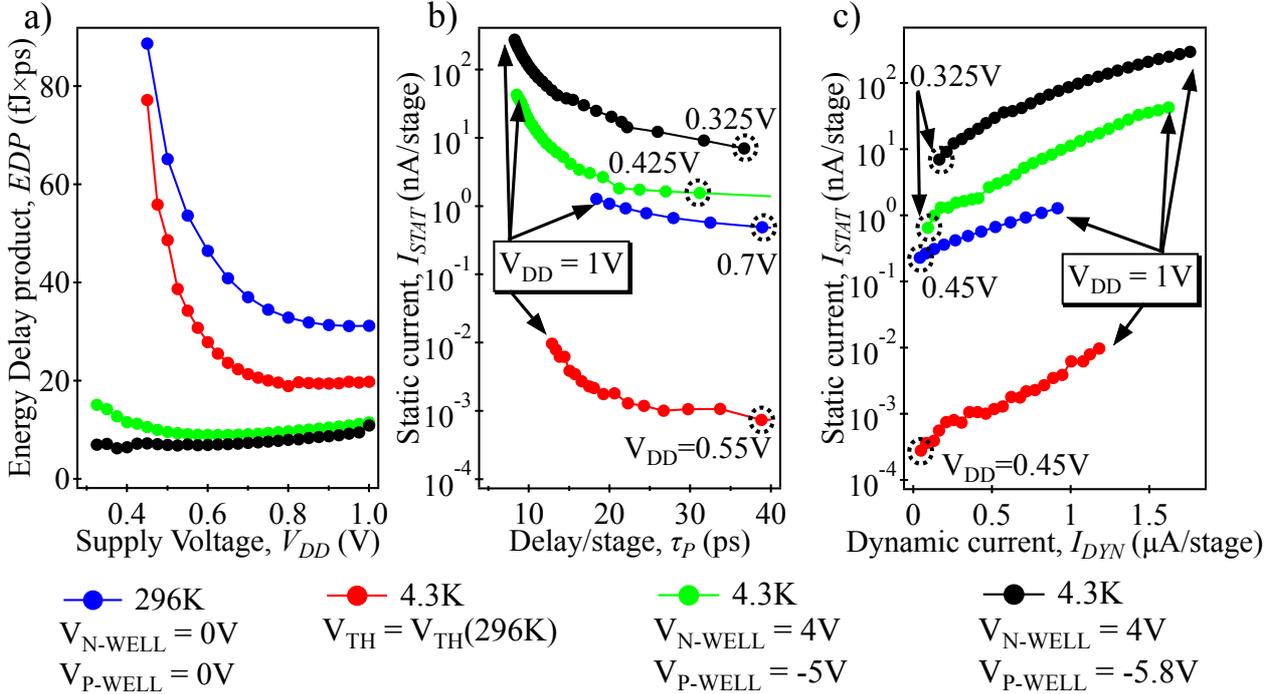

Fig. 8. (a) Comparison of Energy-Delay product vs $V_{DD}$ at room temperature and 4.3K with different forward back-biasing conditions. Note that for the case of very high $V_{FBB}$ ($V_{N-WELL}$=4V, $V_{P-WELL}$=−5.8V) no minimum is observed down to $V_{DD}$=0.325V. (b) Static current vs Delay/stage metric measured at RT and 4.3K for different $V_{DD}$ (initial and final values for the studied $V_{DD}$ interval are given). (c) Static current vs dynamic current for different $V_{DD}$ at 296K and 4.3K. Note that depending on FBB voltage, $I_{DYN}$ between 30nA/stage and 1.76µA/stage can be achieved. Initial and final values for the studied $V_{DD}$ interval are given.

TABLE III
ENERGY EFFICIENCY OPTIMIZATION FOR ULTRA-LOW SUPPLY VOLTAGE UNDER FORWARD BACK-BIASING AT 4.3K

| $V_{DD}$ (V) | $V_{N-WELL}$ (V) / $V_{P-WELL}$ (V) | $\tau_P$ (ps) | $I_{DYN}$ (nA/stage) | $I_{STAT}$ (nA/stage) |
|---|---|---|---|---|
| 0.325 | 4 / -5.8 | 37 | 170 | 7 |
| 0.325 | 4 / -4.8 | 71 | 92 | 0.7 |
| 0.5 | 4 / -5.8 | 17 | 518 | 30 |
| 0.5 | 4 / -4.8 | 21 | 426 | 1.6 |

The data reported in Table III are given to illustrate the tremendous versatility of 28nm FD-SOI technology for both ultra-low power and high performance applications for highly reduced $V_{DD}$ at 4.3K. This highlights the possible trade-offs between the delay and the static/dynamic power dissipation. For instance, in case of $V_{N-WELL}$=4V and $V_{N-WELL}$=−4.8V, we obtain highly improved delay of 31ps with $I_{DYN}$=271nA/stage and $I_{STAT}$=1.56nA/stage if $V_{DD}$ is increased to 0.425V.

## IV.  CONCLUSION

This paper describes, for the first time, electrical characterization of 28nm FD-SOI ring oscillators down to 4.3K. The unique capability of back-biasing in the development of fast power-efficient peripheral circuitry is shown through the analysis of delay per stage, static and dynamic current of the ROs. Also, the trade-off between energy consumption and delay per stage is discussed for a large range of supply voltage ($V_{DD}$ from 0.325 to 1.2V). It is demonstrated that by properly balancing the energy consumption and delay, the maximum benefit in terms of speed and energy consumption can be derived at a very low supply voltage of $V_{DD}$=0.325V only. At 4.3K and high FBB voltage ($V_{N-WELL}$=4V and $V_{P-WELL}$=−5.8V), a very small energy per transition of 0.186fJ with $\tau_P$=37ps and EDP=6.9fJ×ps are achieved at $V_{DD}$=0.325V. These achievements prove that the 28nm FD-SOI platform provides significant opportunities towards optimizing highly-efficient and ultra-low-power cryogenic circuits for large-scale quantum computing. Based on the promising results in this work, other applications such as low temperature sensors, low power neuromorphic circuits [26], or space electronics can be envisioned.

ACKNOWLEDGMENT

This work is partly funded by French Public Authorities (NANO 2017) and Equipex FD-SOI. We also acknowledge financial support from the EU under Project MOS-QUITO (No.688539). The authors would like to thank ST-Crolles characterization team, T. Poiroux, and A. Toffoli for fruitful discussions and help in establishing the measurement protocol for ring oscillators.